\documentclass{article}
\usepackage{wds10,epsf}

\usepackage[square]{natbib}

\lefthead{SHEVTSOV ET AL.}
\righthead{GRAPHENE RIBBON JUNCTIONS}

\begin{document}

\title{Electron transport through symmetrical junctions of semimetal graphene ribbons}

\author{O. Shevtsov}

\affil{National Taras Shevchenko University of Kyiv, Kyiv, Ukraine}

\author{Yu. Klymenko}

\affil{Space Research Institute of NAS and NSA of Ukraine, Kyiv, Ukraine}

%

\begin{abstract}
Based on the nearest-neighbor tight-binding model, we present analytical description of low-energy electron transport through a symmetrical junction of two semimetal armchair ribbons. The results obtained demonstrate the transmission suppression in the vicinity of the neutrality point except for the junction of the ribbons consisting of $3p-1$ and $3p+5$ dimer lines, $p/2$ is odd. Unlike other interconnections, these compounds are shown to be free of even local levels arising at the junction interface and exhibit electron backscattering to be inversely as the square of $p+1$.
\end{abstract}

\begin{article}

\section{Introduction}
Since the discovery of fullerenes, carbon-based materials have been the subject of intense research, which led to the discovery of carbon nanotubes and the fabrication of individual one-atom thick graphene layers. This opens unprecedented avenues for the investigation of quantum transport in low-dimensional 1D and 2D systems, as well as attracting the interest of industries, given the potential for innovative applications. Motivated by possible device applications, the transport properties of graphene ribbons (GRs) has been investigated intensively both in tight-binding \citep{Waka,Roche,KlymShev} and Dirac formalisms \citep{Beenakker,Kats,Blanter}. Both approaches demonstrate that graphene edges drastically change the conducting properties of GRs: zigzag edge
produces localized edge states leading to the metallic type of conductivity, while no localized state appears in an armchair GR\citep{Ando,Brey}. Within the nearest-neighbor tight-binding model GRs can be either metallic if $M=3p-1$ ($p$ is an integer), or semiconducting ($M=3p$ or $M=3p+1$). Here $M$ is a ribbon width in terms of dimer lines.
%

In this paper, based on the nearest-neighbor tight-binding model, we investigate analytically the electron transport through the junctions of armchair graphene ribbons. Specifically, we consider squared symmetrical junctions of two semi-infinite graphene ribbons, which have common longitudinal axis (see Fig. \ref{F2}). Our numerical computations show that there are junctions of a certain type which are almost transparent with respect to the incident wave in the vicinity of zero-energy point (K-point). Unlike them, the low-energy transmission spectrum of other symmetrical junctions demonstrates suppressed transmission. In the present work we find exact analytical expression for the transmission coefficient of the junction near the K-point. As follows from the theory, the low-energy behavior of electron transmission through a semimetal GRJ is  tightly associated with the number ${\cal L}$ of edge states arising at zigzag-shaped boundaries of the ribbons, if they are studied independently. It is important to note, that due to the symmetry of the system, transversal modes of different parity in the leads cannot be mixed. Thus, since the conducting mode in both channels is even, equal number of even edge states in both parts of the junction leads to the practically perfect transmission while a difference between them results in the transmission suppression. In the latter case, a zero-energy localized state appears at the interface between two ribbons. Thus, in agreement with \citep{Zheng_1,Zheng_2}, the low-energy transmission suppression arises from the local level appearance near the junction interface. If a junction is free of the even local levels, the transmission exhibits practically perfect propagation.

\section{Wave solution for electron in armchair GRs}

\begin{figure}[htb]
\begin{center}
\begin{tabular}{c}
  \epsfxsize=120mm
  \epsfbox{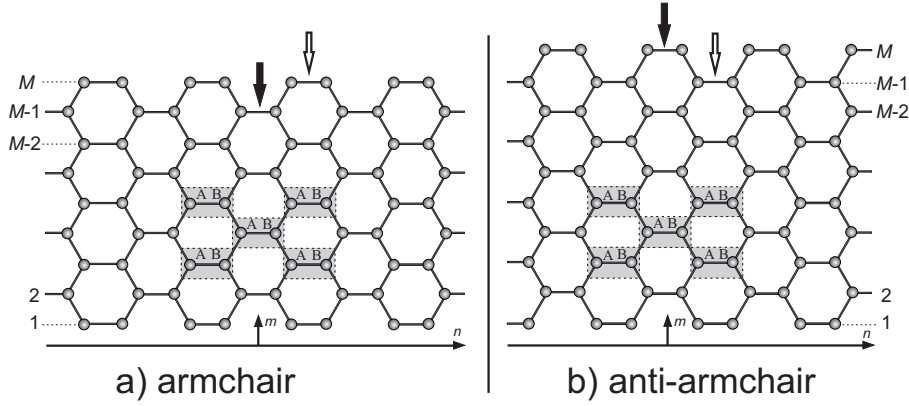}
\end{tabular}
\end{center}
\caption{Sketch of infinite graphene ribbons in armchair configuration (left panel) and in anti-armchair configuration (right panel), consisting of $M$ dimer lines in the transverse direction. $M$ is even for anti-armchair configuration and odd for armchair one. Black or empty arrows denote cells with integer $n$, see details in text.
Shadowed blocks show the nearest neighboring cells.}\label{F1}
\end{figure}

The infinite armchair GR is considered as a set of
elementary cells of two atoms
$\alpha={\rm A,B}$ with edges in armchair and in anti-armchair configurations, see Fig.
\ref{F1}. We enumerate the cells by two numbers $\{n,m\}$ possessing  both integer and half-integer values simultaneously, $-\infty<n<\infty$.
By taking the tight-binding representation for
molecular orbitals we come to the set of linear Schr\"{o}dinger
equations
\begin{equation}\label{A1}
\left\{\!\!
\begin{array}{l}
-E\psi_{n,m,{\rm A }}=\psi_{n,m,{\rm B
}}+\psi_{n-\frac{1}{2},m+\frac{1}{2},{\rm B
}}+\psi_{n-\frac{1}{2},m-\frac{1}{2},{\rm B }},\\ -E\psi_{n,m,{\rm
B}}=\psi_{n,m,{\rm A}}+\psi_{n+\frac{1}{2},m+\frac{1}{2},{\rm A
}}+\psi_{n+\frac{1}{2},m-\frac{1}{2},{\rm A}}
\end{array}
\right.
\end{equation}
with respect to wave function components taken at $2{\rm p}_z$
orbital of $\alpha$-th atom in the elementary cell,
$E\equiv E/|\beta|$ is the electron energy in units of hopping
integral $|\beta|$ between the nearest-neighbor carbon atoms,
$\beta<0$. Carbon site-energy is assumed to be zero and serves as
the reference.

Since boundary carbons of infinite armchair GRs have only two
neighbors (not three as all inner atoms), it demands the wave function to vanish on the set of absent sites closest to the boundary atoms \citep{Robinson,Zheng,Akhmerov}. Evidently, analytical form of the boundary conditions depends on a numeration method of the elementary cells along $n$ direction. If index $n$ takes integer value in the regions marked by the black arrows on  Fig. \ref{F1}, we get $\psi_{n,0,\alpha}=0$ for the lower boundaries, and $\psi_{n,N+1,\alpha}=0$, $\psi_{n+1/2,N+1/2,\alpha}=0$ for the upper boundaries in armchair (Fig. \ref{F1}a) and anti-armchair (Fig. \ref{F1}b) configurations, respectively. Here $N=[M/2]$ is the number of elementary cells in the marked regions, $M$ is the number of dimer lines in the transverse direction of the armchair ribbon, and $[M/2]$ is the integer part of $M/2$. Similarly, if we set $n$ to be an integer in the regions marked by the empty arrows, one can obtain $\psi_{n+1/2,1/2,\alpha}=0$ for the lower boundaries, and $\psi_{n+1/2,N+1/2,\alpha}=0$, $\psi_{n,N+1,\alpha}=0$ for the upper ribbon boundaries on the corresponding Figures. Here $N=[M/2]+1$ for armchair configuration and $N=[M/2]$ for anti-armchair one.

The nontrivial solutions to Eq.(\ref{A1}) satisfying the appropriate boundary conditions can be written as follows
\begin{equation}\label{A2}
\psi_{n,m,\alpha}(j)=\frac{\chi_j(m)}{\sqrt{M+1}}\,e^{ik_j n}\left\{\!\!
\begin{array}{ll}
1,&\alpha={\rm A}, \\ e^{i\theta_j},&\alpha={\rm B}.
\end{array}
\right.
\end{equation}
Here we introduced the transversal wave functions for the numeration methods marked by black and empty arrows respectively
\begin{equation}\label{A3}
\chi_j(m)=\sin(\xi_jm),\qquad \chi_j(m)=\sin[\xi_j(m-1/2)],
\end{equation}
where quantity
\begin{equation}\label{A4}
\xi_j=\frac{2\pi j}{M+1}, \qquad 1\leq j \leq \left[\frac{M}{2}\right]
\end{equation}
has meaning of the dimensionless transversal wave number for electron in armchair GRs.
The longitudinal wave number $k_j$ from Eq. (\ref{A2}) can be found from dispersion relation \citep{Ando}, which can be derived by plugging in Eq. (\ref{A2}) to Eq. (\ref{A1}) and demanding zero determinant. Thus,
\begin{equation}\label{A5}
\cos\frac{k_j}{2}=\frac{E^2-1-4\cos^2\frac{\xi_j}{2}}{4\cos\frac{\xi_j}{2}},
\end{equation}
and quantity $\theta_j$ is defined by expression
\begin{equation}\label{A6}
e^{i\theta_j}=-\frac{1+2\cos\frac{\xi_j}{2}\,e^{ik_j/2}}{E}.
\end{equation}

Given $E$ and $\xi_j$, real-valued number $k_j$ determines a propagating state in the channel. Here real-valued quantity $\theta_j$ has meaning of phase shift between A and B atoms in the elementary cell \citep{KlymShev} and state (\ref{A2}) is normalized on a unit cell\footnote{By definition, the unit cell of the ribbon is a union of elementary cells along the translation period of the ribbon. It consists of all $\{n,m\}$ and $\{n+1/2,m\}$ elementary cells, where the number $n$ is fixed.} of the ribbon. The group velocity connected with the propagating state can be written as follows
$$
v_j=\frac{dE}{dk}=-\frac{1}{E}\cos\frac{\xi_j}{2}\sin\frac{k_j}{2}=\frac{\sin\theta_j}{2}.
$$
Thus, the propagating wave (\ref{A2}) runs from left to right in the valence subbands and vice versa for the conducting ones.

It should be noted that the choice of the elementary cell numeration along $n$ direction of the infinite graphene ribbons remains arbitrary. Obviously, the value of the wave function component $\psi_{n,m,\alpha}$ does not depend on certain choice. Considering armchair GR junctions, we'll refer elementary cells $n=0$ to the junction interface. In this case, the numeration of elementary cells in the ribbons becomes unique.

\section{General equations for scattering amplitudes}

\begin{figure}[htb]
\begin{center}
\begin{tabular}{c}
  \epsfxsize=120mm
  \epsfbox{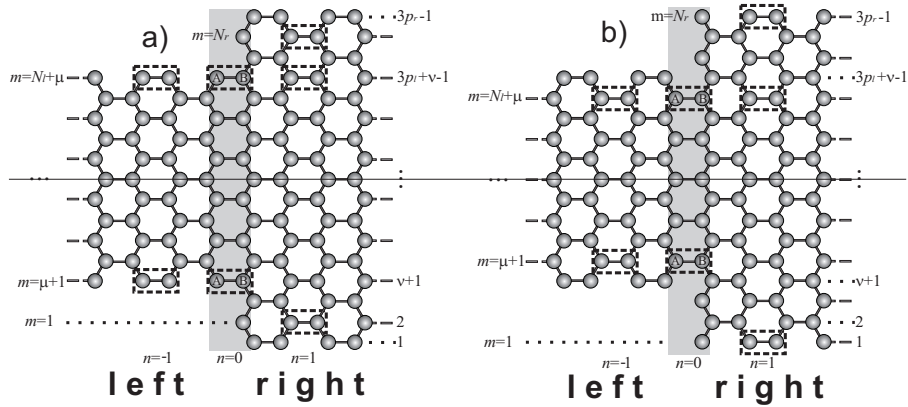}
\end{tabular}
\end{center}
\caption{Two possible geometries of the symmetrical junction of armchair GRs. Shadowed rectangle depicts the scattering region. Transition from case a to case b is provided by deletion of one zigzag line from the shoulders, and vice versa. Figure represents junctions with three dimer lines long shoulders as they are the subject of the main text.}\label{F2}
\end{figure}

Now we consider a symmetrical junction of two semi-infinite graphene ribbons with armchair edges, which have common longitudinal axis (see Fig. \ref{F2}). Such junctions can be, obviously, constructed only of ribbons in armchair configuration. The corresponding matching conditions are as follows
\begin{equation}\label{A7}
\psi^{l}_{0,m,B}=\psi^{r}_{0,m,B}\,,\quad \mu+1\leq m\leq N_{l}+\mu\,,
\end{equation}
\begin{equation}\label{A8}
\psi^{r}_{0,m,A}=\left\{\!\!
\begin{array}{l}
0\,,\qquad\quad\;1\leq m\leq \mu,\\
\psi^{l}_{0,m,A}\,,\quad \mu+1\leq m\leq N_{l}+\mu\,,\\
0\,,\qquad\quad\;N_{l}+\mu+1\leq m\leq N_{r}.
\end{array}
\right.
\end{equation}
Later on, by indexing $\nu=l,r$ we will distinguish quantities referring to the corresponding left ($l$) and right
($r$) leads.

The incident wave with the transverse mode $j_0$ in the left lead, meeting the interface between two ribbons, scatters into both propagating and evanescent modes. The general wave solutions for the electron in both leads should be constructed as a linear combination of expression (\ref{A2}) over all transverse modes $j$ in the corresponding channel. In particular, for $E<0$ we obtain
\begin{equation}\label{A9}
\psi^{l}_{n,m,\alpha}=\psi^{l}_{n,m,\alpha}(j_0)+\sum_{j=1}^{[M_{l}/2]}r_{j,j_0}\psi^{l*}_{n,m,\alpha}(j),\qquad \psi^{r}_{n,m,\alpha}=\sum_{j=1}^{[M_{r}/2]}t_{j,j_0}\psi^{r}_{n,m,\alpha}(j).
\end{equation}
Here co-factors $r_{j,j_0}$ and $t_{j,j_0}$ for the propagating  mode $j$ have meaning of the scattering amplitudes in the corresponding channel. When $k^{\nu}_j$ is complex, the state $\psi^{\nu}_{n,m,\alpha}(j)$ describes an evanescent eigenmode in the corresponding lead. In this case $r_{j,j_0}$ and $t_{j,j_0}$ are unphysical quantities and there is no need to normalize $\psi^{\nu}_{n,m,\alpha}(j)$.

Plugging in expressions (\ref{A9}) into conditions (\ref{A7}), (\ref{A8}) and
taking into account orthogonality of functions $\{\chi^{\rm \nu}_j(m)\}$, we come to a set of equations
\begin{equation}\label{A10}
\begin{array}{l}
\displaystyle e^{i\theta^{l}_j}\delta_{j,j_0}+e^{-i\theta^{l}_j}r_{j,j_0}=\sum_{j^{\prime}=1}^{[M_r/2]}g_{j,j^{\prime}}e^{i\theta^{r}_{j^{\prime}}}t_{j^{\prime},j_0}\,,\\
\displaystyle t_{j,j_0}=\sum_{j^{\prime}=1}^{[M_l/2]}g_{j^{\prime},j}\left[\delta_{j^{\prime},j_0}+r_{j^{\prime},j_0}\right]\,,
\end{array}
\end{equation}
where $\delta_{j,j_0}$ is the Kronecker delta and
\begin{equation}\label{A11}
g_{j,j^{\prime}}=\frac{4}{\sqrt{(M_{l}+1)(M_{r}+1)}}\sum_{m=1}^{N_{l}}\chi^{l}_{j}(m)\chi^{r}_{j'}(m+\mu)\,
\end{equation}
is a matrix coupling transverse modes of the left and right leads.

Eqs. (\ref{A10}) hold for arbitrary types of ribbons (semimetal or semiconducting) forming a squared junction with a common edge. Solving them with respect to scattering amplitudes $t_{j,j_0}$, one can obtain the transmission probabilities $T_{j,j_0}=(v^r_j/v^l_{j_0})|t_{j,j_0}|^2$ and the transmission spectrum $T(E)=\sum_{j,j_0}T_{j,j_0}(E)$ of the  junction.
\begin{figure}[htb]
\begin{center}
\begin{tabular}{c}
  \epsfxsize=120mm
  \epsfbox{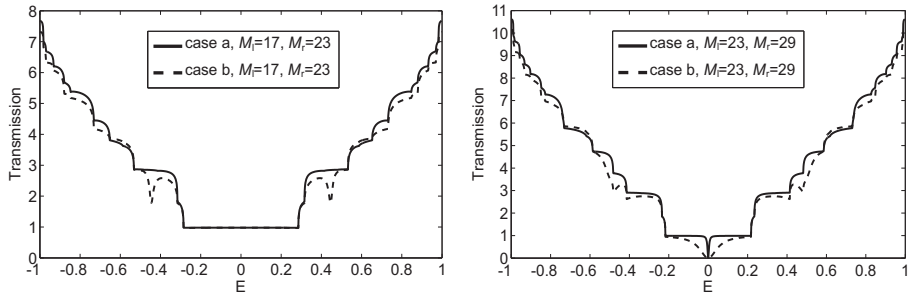}
\end{tabular}
\end{center}
\caption{Transmission spectrum for semimetal $17/23$ and $23/29$ graphene ribbon junctions. The corresponding junction geometry is specified on  Fig. \ref{F2}, where cases a and b refer to the left and right panels correspondingly.}\label{F3}
\end{figure}

Later on, we focus on the metallic graphene junction $M_l/M_r$ where $M_{\nu}=3p_{\nu}-1$, and $p_{\nu}$ is an integer. Numerical modeling for $17/23$ junctions, the left panel on Fig. \ref{F3}, demonstrates the nonzero transmission, however, calculations carried out for $23/29$ junctions show transmission suppression in the vicinity of zero-energy point. Thus, we can find that junction family $(3p-1)/(3p+5)$ with odd $p/2$ demonstrates practically perfect transmission while the transmission is suppressed for even $p/2$. More detailed analysis of transmission properties of semimetal graphene junctions observed at $E=0$ is the subject of the next section.

\section{Propagation at the neutrality point in semimetal junctions}

Detailed analysis of band structure based on  Eq. (\ref{A5}) shows that in the vicinity of zero-energy point ($E=0$), the only transversal mode available for propagating has number $j_{\nu}=p_{\nu}$ in each channel. Meantime all other modes are vanishing since in that energy region Eq. (\ref{A5}) does not give real-valued longitudinal component of the wave vector, $k_j^{\nu}$. By substitution of $\xi^{\nu}_{p_{\nu}}=2\pi/3$  to energy dispersion (\ref{A5}) we obtain $k^{\nu}_{p_{\nu}}=2\pi$. To find $\theta^{\nu}_{p_{\nu}}$ we plug in the energy dispersion to expression for $\cos \theta^{\nu}_{p_{\nu}}$ following from Eq. (\ref{A6}), $\theta^{\nu}_{p_{\nu}}=\pi/2$. Thus, one can write down the following expression for the propagating waves in the leads
\begin{equation}\label{A12}
\psi^{\nu}_{n,m,\alpha}(p_{\nu})=\frac{\chi^{\nu}_{p_{\nu}}}{\sqrt{M_{\nu}+1}}
\textstyle\left\{\!\!
\begin{array}{ll}
1,&\alpha={\rm A}, \\ i,&\alpha={\rm B},
\end{array}\right.\quad \nu={l,r}.
\end{equation}

Evanescent electron state in the leads corresponds to a complex-valued $k_j^{\nu}$. It can also be constructed from state (\ref{A2}) by substituting $k_j^{\nu}/2=\pi+i\delta_j^{\nu}/2$ into definition (\ref{A5})  taken for $E=0$,
\begin{equation}\label{A13}
e^{\delta_{j}^{\nu}/2}=\left\{\!\!
\begin{array}{ll}
2\cos\frac{\xi^{\nu}_{j}}{2},&2\cos\frac{\xi^{\nu}_{j}}{2}>1\Leftrightarrow j<p_{\nu}, \\
1/[2\cos\frac{\xi^{\nu}_{j}}{2}],&2\cos\frac{\xi^{\nu}_{j}}{2}<1\Leftrightarrow p_{\nu}<j\leq \left[\frac{M_{\nu}}{2}\right].
\end{array}
\right.
\end{equation}
With help of Eqs. (\ref{A6}) and (\ref{A13}) we must choose regular evanescent solutions for both leads.
Finally,
\begin{equation}\label{A14}
\begin{array}{ll}
\psi_{n,m,\alpha}^{l}(j>p_{l})=\frac{\chi^{l}_{j}(m)}{\sqrt{M_l+1}}\,e^{\delta_{j}^{l} n}
\left\{\!\!
\begin{array}{ll}
1,&\alpha={\rm A}, \\
0,&\alpha={\rm B},
\end{array}
\right.&
\psi_{n,m,\alpha}^{r}(j>p_{r})=\frac{\chi^{r}_{j}(m)}{\sqrt{M_r+1}}\,e^{-\delta_{j}^{r} n}
\left\{\!\!
\begin{array}{ll}
0,&\alpha={\rm A}, \\
1,&\alpha={\rm B},
\end{array}
\right.\\
\psi_{n,m,\alpha}^{l}(j<p_{l})=\frac{\chi^{l}_{j}(m)}{\sqrt{M_l+1}}\,e^{\delta_{j}^{l} n}
\left\{\!\!
\begin{array}{ll}
0,&\alpha={\rm A}, \\
1,&\alpha={\rm B},
\end{array}
\right.&
\psi_{n,m,\alpha}^{r}(j<p_{r})=\frac{\chi^{r}_{j}(m)}{\sqrt{M_r+1}}\,e^{-\delta_{j}^{r} n}
\left\{\!\!
\begin{array}{ll}
1,&\alpha={\rm A}, \\
0,&\alpha={\rm B}.
\end{array}
\right.
\end{array}
\end{equation}
It is important to note that the evanescent eigenmodes $j>p_{\nu}$ in Eqs. (\ref{A14}) coincide identically with the edge state solutions arising at zigzag-shaped boundaries of the corresponding isolated semi-infinite leads where the appropriate boundary conditions $\psi_{0,m,{\rm B}}^{l}(j>p_{l})=0$, $\psi_{0,m,{\rm A}}^{r}(j>p_{r})=0$ must be satisfied \citep{Waka,Nakada}. The number of such modes (with $j>p_{\nu}$) in the corresponding ribbon is determined by dependence
$$
{\cal L}_{\nu}=\left[\frac{M_{\nu}}{2}\right]-p_{\nu}=\left[\frac{p_{\nu}-1}{2}\right],
$$
since $1\leq j\leq \left[\frac{M_{\nu}}{2}\right]$.

To obtain equations for scattering amplitudes, one needs to satisfy interface conditions (\ref{A7}) and (\ref{A8}), using the linear combination of solutions (\ref{A12}) and (\ref{A14}). The result reads,
\begin{equation}\label{A15}
\begin{array}{l}
\displaystyle i\delta_{j,p_l}(2-\bar{r}_{j,p_l})+\bar{r}_{j,p_l}\eta(p_l-j)=
\sum_{j'=1}^{p_r+{\cal L}_{r}}g_{j,j'}t_{j',p_l}[i\delta_{j'\!,p_r}+\eta(j'-p_r)],\\
\displaystyle t_{j,p_l}[\delta_{j,p_r}+\eta(p_r-j)]=
\sum_{j'=1}^{p_l+{\cal L}_{l}}g_{j'\!,j}[\delta_{j'\!,p_l}\bar{r}_{j,p_l}+\bar{r}_{j'\!,p_l}\eta(j'-p_l)].
\end{array}
\end{equation}
Here $\bar{r}_{j,p_l}=r_{j,p_l}+\delta_{j,p_l}$, $\eta(x)=1$ if $x>0$ and $\eta(x)=0$ otherwise. Using Eq. (\ref{A3}) and definition (\ref{A4}), one can find the following properties of the coupling matrix $g_{j,j'}$. First, due to the symmetry of the system, transversal modes of different parity cannot be mixed, which means that $g_{j,j'}\neq0$ only if $j$ and $j'$ are odd or even simultaneously. Second, when $j'=p_r$, one can verify that
\begin{equation}\label{A17}
g_{j,p_r}=g\delta_{j,p_l}, \quad g=\sqrt{p_l/(p_l+2)}.
\end{equation}
Since the propagating mode in both leads is even, one must solve Eqs. (\ref{A15}) taking into account only even-numbered modes. This is the subject of the next section.

\section{Analytical solution for transmission probability}\label{5}

First we introduce the number of even edge states in the leads, $\mathcal{L}^{\rm even}_{\nu}=[\mathcal{L}_{\nu}/2]$. Due to relation (\ref{A17}), the second equation in (\ref{A15}) taken for $j\geq p_r$ is simplified to the set of linearly independent  equations
\begin{equation}\label{A18}
t_{p_r,p_l}-g\bar{r}_{p_l,p_l}=0,
\end{equation}
\begin{equation}\label{A19}
g_{p_l,j}\,\bar{r}_{p_l,p_l}+\displaystyle\sum_{j'-p_l=2,4,\dots}^{{\cal L}^{\rm even}_{l}}g_{j',j}\bar{r}_{j',p_l}=0,\quad j-p_r=2,4,\dots,{\cal L}^{\rm even}_{r}.
\end{equation}
If $\mathcal{L}^{\rm even}_{r}\geq\mathcal{L}^{\rm even}_{l}+1$, homogeneous system (\ref{A19}) always has trivial solution $\bar{r}_{j\geq p_l,p_l}=0$, which leads to the transmission suppression $T=|t_{p_r,p_l}|=0$. If $\mathcal{L}^{\rm even}_{r}=\mathcal{L}^{\rm even}_{l}$, then the first equation in (\ref{A15}) taken for even $j>p_l$ leads to the closed set of homogeneous equations. That system has trivial solutions $t_{j>p_r,p_l}=0$. Substitution of this result to the first equation in (\ref{A15}) taken for $j=p_l$ gives the equation
\begin{equation}\label{A20}
gt_{p_r,p_l}+\bar{r}_{p_l,p_l}=2.
\end{equation}
Solving Eqs. (\ref{A18}) and (\ref{A20}) with respect to the transmission amplitude $t_{p_r,p_l}$, one can obtain the following expression for transmission coefficient
\begin{equation}\label{A21}
T=|t_{p_r,p_l}|^2=\frac{4g^2}{(1+g^2)^2}=1-\frac{(p_r-p_l)^2}{(p_r+p_l)^2}.
\end{equation}
Condition $\mathcal{L}^{\rm even}_{r}=\mathcal{L}^{\rm even}_{l}$ holds only when $p_r=p_l+2$ and $p_l/2$ is odd. Thus, we obtain
\begin{equation}\label{A22}
T=1-(p_l+1)^{-2}.
\end{equation}

There is a physical reason for such behavior of the transmission coefficient. Forming a junction between two ribbons may lead to local level occurrence near the junction interface. To obtain the condition for existence of the states localized at the junction interface one should use only linear combination of evanescent solutions (\ref{A14}) omitting the propagating ones. Utilizing boundary conditions (\ref{A7}) and (\ref{A8}) one can find the linear homogeneous system of equations. It has nontrivial solution only when ${\cal L}^{\rm even}_{l}\ne{\cal L}^{\rm even}_{r}$. Thus, transparent junctions mentioned above are free of even local levels which causes good conducting properties of such systems. In contrast to them other junctions have zero-energy local level of degeneracy ${\cal D}={\cal L}^{\rm even}_{r}-{\cal L}^{\rm even}_{l}$ which leads to the full suppression of transmission in the vicinity of $E=0$.

\section{Conclusions}
The results obtained in the present work show that the low-energy propagation through a symmetrical graphene junction is directly connected to existence of even localized states near the junction interface. If junctions are free of such local levels, they exhibit almost perfect propagation described by dependence (\ref{A22}). Other GRJs possess local level suppressing the propagation in the vicinity of the K-point.

Symmetrical junctions do not exhaust all types of transparent junctions. There are also common-edged $(3p-1)/(3p+2)$ (p is odd) junctions with three dimer lines long shoulder which are free of local levels too. Applying the same technique as for symmetrical junctions one can find that for common-edged junctions the reflection coefficient turns out to be inversely as the square of $(2p_l+1)$ \citep{KlymShev1}. Our calculations performed with help of Eq. (\ref{A10}) and Eq. (\ref{A11}) for such systems (specifically, $20/23$ and $23/26$ junctions) are in perfect agreement with those obtained numerically in the framework of Landauer-Buttiker formalism \citep{Zheng_1,Zheng_2}. The calculations made show that common-edged junctions with upward-shifting the narrower lead on one or two dimer lines conserves the unsuppressed transmission, though the solution for the transmission coefficient cannot be found analytically.

Obtained results give an opportunity to go further and consider more complicated systems, such as: quantum dot structure, multiple quantum dot structures, superlattices, constrictions of armchair graphene ribbons, junction of armchair graphene ribbon and zigzag nanotube, etc.

\acknowledgments 
{The present work was supported by STCU Grant \#\,21-4930/08.}

\bibliographystyle{egs}
\bibliography{KlymShev}       

\end{article}
\end{document}